\documentclass[sigconf]{acmart}

\usepackage{booktabs}
\usepackage[normalem]{ulem}  % for \sout => should be removed later
\usepackage{bbold}  % for \mathbb
\usepackage{subfigure}  % for \subfigure 
\usepackage{placeins}  % for \FloatBarrier
\usepackage{multirow} % for \multirow
\usepackage{moreverb} % for verbatimtab
\usepackage{enumerate} % for \begin{enumerate}
\usepackage{enumitem}  % for setting margins on itemize
\usepackage{tabularx} % for \tabularx
\usepackage{pifont}
\usepackage{xspace}

\newcolumntype{L}[1]{>{\raggedright\arraybackslash}p{#1}}
\newcolumntype{C}[1]{>{\centering\arraybackslash}p{#1}}
\newcolumntype{R}[1]{>{\raggedleft\arraybackslash}p{#1}}

\newcommand{\firstDataset}{Dataset 1\xspace}
\newcommand{\secondDataset}{Dataset 2\xspace}

\newcommand{\simpleArch}{NeuType1\xspace}
\newcommand{\complexArch}{NeuType2\xspace}

\newcommand{\inputA}{$\vect{A}$\xspace} 
\newcommand{\inputB}{$\vect{B}$\xspace} 
\newcommand{\inputC}{$\vect{C}$\xspace} 

\newcommand{\metric}{NDCG@1\xspace}

\newcommand{\highSigBASE}{$^\ddag$}  % sig. at 0.001; vs baseline
\newcommand{\sigARCHS}{$^\Box$}  % sig. at 0.05; arch. C vs arch. S
\newcommand{\highSigARCHS}{$^\Diamond$}  % sig. at 0.001; arch. C vs arch. S

\newcommand{\uri}[1]{\scriptsize\texttt{#1}\normalsize}

\newcommand{\vect}[1]{\boldsymbol{#1}}

\begin{document}

\title{NeuType: A Simple and Effective Neural Network Approach for Predicting Missing Entity Type Information in Knowledge Bases}

%\author{~\affiliation{~}{~}}

% -------------
% Authors
\author{Jon Arne B{\o} Hovda}
\affiliation{University of Stavanger}
\email{ja.hovda@stud.uis.no}
\author{Dar\'{i}o Garigliotti}
\affiliation{University of Stavanger}
\email{dario.garigliotti@uis.no}
\author{Krisztian Balog}
\affiliation{University of Stavanger}
\email{krisztian.balog@uis.no}
%
% -------------

\begin{abstract}
Knowledge bases store information about the semantic types of entities, which can be utilized in a range of information access tasks.  This information, however, is often incomplete, due to new entities emerging on a daily basis.  We address the task of automatically assigning types to entities in a knowledge base from a type taxonomy.  
Specifically, we present two neural network architectures, which take short entity descriptions and, optionally, information about related entities as input.  
Using the DBpedia knowledge base for experimental evaluation, we demonstrate that these simple architectures yield significant improvements over the current state of the art.  

\end{abstract}

%\ccsdesc[500]{Information systems~Information extraction}

\keywords{Entity types; neural entity type prediction; knowledge bases}

\maketitle

\section{Introduction}
\label{sec:intro}

% ------- Types: importance and challenges
A characteristic property of a given entity in a knowledge base (KB) is its \emph{type}.  
Entity types, as semantic classes grouping several entities, are a key information signal, which can be exploited in a variety of tasks in information extraction, document classification, natural language processing, and information retrieval~\citep{Balog:2018:EOS,Garigliotti:2019:IET}.  
However, the type information associated with entities in the KB is often incomplete, imperfect, or missing altogether for some entities.  
In addition, new entities emerging on a daily basis also need to be mapped to one or more types of the underlying type system.  
For alleviating these problems, in this paper we address the challenge of predicting missing type information for entities in a KB.  

% ------- SDType: pros and cons
SDType~\citep{PAULHEIM:2013:TIO} is a state-of-the-art entity type prediction method that leverages links between entities and properties in order to support KBs with incompleteness and noisy labels.
SDType, however, requires substantial amounts of information from the KB, in order to obtain the statistical distributions of links and properties.  
Our goal is therefore to enable typing entities based on limited information, that is, based solely on entity descriptions.   
This is particularly important for emerging entities, which usually have only a brief description and lack most of the additional structured knowledge.  At the same time, if information about related entities is available, we wish to be able to leverage that as well.

% ------- THIS paper: overview, contributions
We propose two simple fully-connected feedforward neural network (FNN) architectures, and consider different ways to represent an input entity in order to predict a single type label.  
We create two test collections based on DBpedia (ver. 2016-10), one focusing on established entities and another focusing on emerging entities.  
Our results show that these simple FNN architectures are able to provide a substantial and significant improvement over the current state of the art.  
Furthermore, type prediction results based solely on short entity descriptions can significantly be improved when incorporating information about related entities.
We also observe that the deeper FNN performs better on established entities, whereas the shallower architecture is sufficient to predict types for emerging entities that have only short descriptions.  
  % intro
\section{Related Work}
\label{sec:rel}

% ------- Auto. entity typing: related tasks and challenges
The problem of automatic entity typing has been studied under several related tasks, including named entity recognition, entity linking, and type inference~\citep{Balog:2018:EOS}.  
Regarding the challenges of this problem, \citet{GANGEMI:2012:ATO} distinguish between \emph{extensional} coverage, i.e., the number of typed resources, and \emph{intensional} coverage, i.e., conceptual completeness.  
They introduce Tipalo, a tool that makes use of the natural language definition (i.e., abstract) of an entity in Wikipedia.  
We also exploit the textual description of an entity, but whereas Tipalo infers types from graph patterns over logical interpretations of the entity definitions, we model type label assignment as a deep supervised learning task.  

%------ Related approaches in fine-grained typing in context
Similar approaches are used in the related task of fine-grained entity typing in context.  
\citet{LIN:2012:NNP} exploit entity definitions to map entities into Freebase types by analyzing n-grams in the textual relations around entity mentions.  
\citet{NAKASHOLE:2013:FGS} address typing emerging entities, that are of  particular importance for informative knowledge, from news streams and social media.     
Multi-instance, multi-label typing algorithms are used in \citep{YAGHOOBZADEH:2017:NMF} over KB data and annotated contexts of entities in a corpus.  
Rather than the more extensive evidence sources that these approaches exploit in the context of entity occurrences, we rely on short definition-like descriptions as our input.  

% ------- SVMs hierarchy vs single model
\citet{KLIEGR:2016:L2A} predict the entity type by linearly combining the output distributions from several techniques: (i) string matching and statistical inference on an external hypernyms dataset extracted from Wikipedia, and (ii) hierarchies of classifiers trained on entity short abstracts and on Wikipedia categories.  
Unlike their approach, that requires multiple components and training several classifiers, we propose a single end-to-end model.   

% ------- Another MLP
Corpus-level entity typing is also used for knowledge base completion.  
A multilayer perceptron approach has been proposed using word embeddings 
\citep{YAGHOOBZADEH:2015:CFE}.  
While similar to this underlying approach, we employ a larger type system (112 FIGER types vs. 760 DBpedia types), and utilize various input entity representations.

% ------- SDType
\subsection{SDType}
\label{subsec:sdtype}
%\subsectionshrink

SDType, presented in \citep{PAULHEIM:2013:TIO}, and further expanded upon in \citep{PAULHEIM:2014:ITQ}, utilizes links between instances in a KB to infer types using a weighted voting approach.  
The main assumption is that some relationships between entities only occur with particular entity types.  
As an example, from the statement $x$ \texttt{dbo:location} $y$, it can infer with high confidence that $y$ is of type \texttt{Place}.  
Unlike other type prediction methods, SDType can be implemented on virtually any cross-domain dataset \cite{PAULHEIM:2013:TIO}.  
Its evaluation on DBpedia 3.8 reports an F-measure of 0.885, making SDType outperform all the compared methods, including Tipalo~\citep{GANGEMI:2012:ATO}, which achieves an F-measure of 0.75.
Since DBpedia 3.9, the type assignments obtained by SDType, available for a large subset of entities, are distributed with DBpedia.
While both SDType and our approaches leverage entity relationship information, an important difference is that SDType requires typed links, whereas our approaches consider the mere presence of a relationship, i.e., links are non-typed.  
This significantly weakens the requirements for an input entity to be typed.

  % rel
\section{Approach}
\label{sec:approach}

We begin by describing the overview of our proposed model architectures, and then provide the details of their input components.

% -------
\subsection{Architecture Design}
\label{subsec:architecturedesign}

% --- MLP overview
Our approach is based on a multilayer perceptron, a simple neural network architecture, with vector representations of entities as inputs and a softmax operation on the output layer, to obtain a probability distribution among all types.  
This model is simple yet also flexible to account for combining various input representations, possibly of different dimensions, as shown in \cite{YAGHOOBZADEH:2015:CFE}, where a similar architecture is used for fine-grained typing of entity mentions.  % in text.

\begin{figure}[t]
	\vspace*{-0.75\baselineskip}
    \centering
		\subfigure[\simpleArch]{
	        \includegraphics[width=0.38\textwidth]{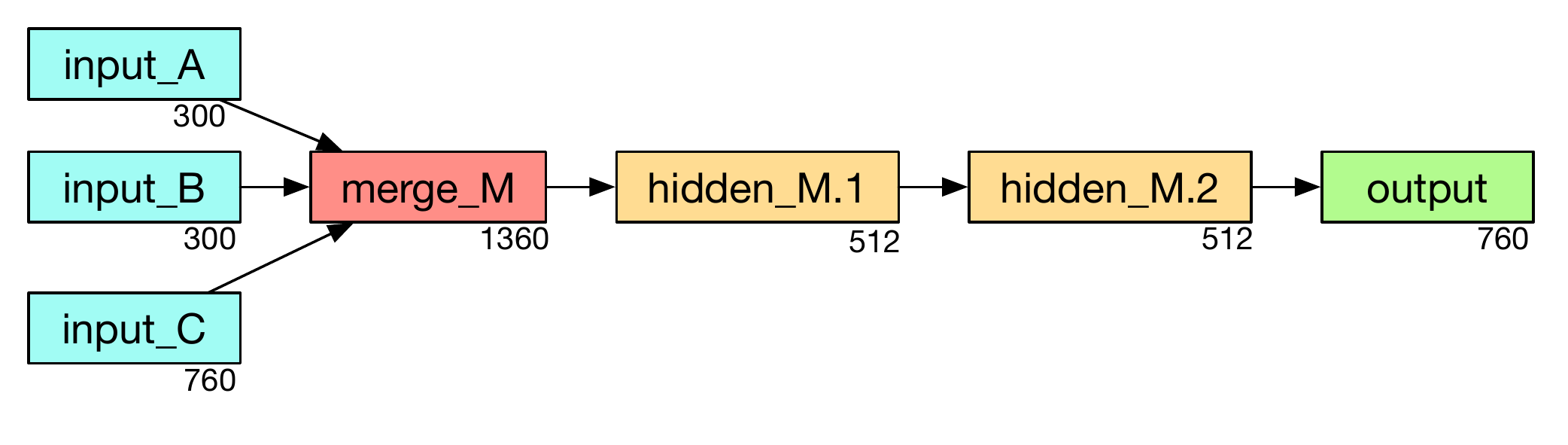}
			\label{fig:neuralnetsimple}
		}
		\hspace{0.2in}
		\subfigure[\complexArch]{
	        \includegraphics[width=0.475\textwidth]{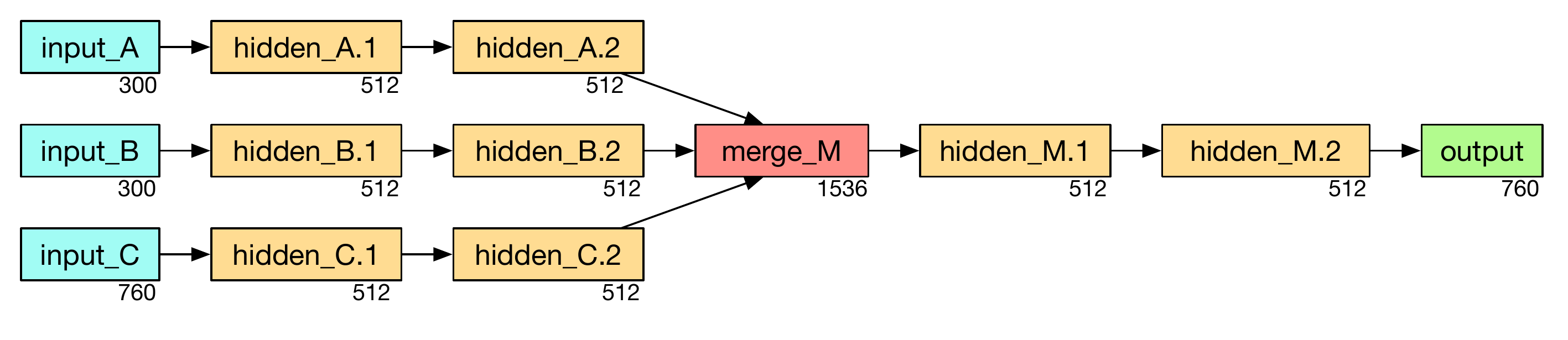}
			\label{fig:neuralnetcomplex}
		}
	\centering
	\vspace*{-0.85\baselineskip}
	\caption{Neural architectures.  Arrows indicate fully-connected layers; subscripted, number of nodes in a layer.}
	\label{fig:modes}
	\vspace*{-1.1\baselineskip}
\end{figure}

% --- Architecture NeuType 1
Figure~\ref{fig:neuralnetsimple} presents our first architecture, \simpleArch.  
It consists of a fully-connected feedforward neural network, and is able to handle different entity vector representations, which are given by input components \texttt{input\_A}, \texttt{input\_B}, and \texttt{input\_C} (cf.~Sect.~\ref{subsec:inputcomponents}).  
A merge layer \texttt{merge\_M} concatenates the available inputs into hidden layers \texttt{hidden\_M.1} and \texttt{hidden\_M.2}.  
The outputs are transformed by softmax into a probability distribution across all possible 760 type labels in the DBpedia Ontology (we discard \texttt{<owl:Thing>}).  %\footnote{We discard the \texttt{<owl:Thing>} root type, which is meaningless for our task.}  
This model resembles a simple learning framework, where a neural classifier is applied on a merging of multiple input vectors~\citep{Le:2014:DRS}.

% --- Architecture NeuType 2
Unlike in \simpleArch, in \complexArch, depicted in Fig.~\ref{fig:neuralnetcomplex}, each input component is firstly fully connected to its own stack of hidden layers.  
In this way, its depth allows it to better capture each input entity representation, before combining them by vector concatenation.  
Similar deep merging networks have proven to be effective versus another textual input compositions for classification tasks~\citep{Iyyer:2015:DUC}.  

% --- Classif. task
When defining the model output, we are interested in finding a single (most correct) type.  
We therefore approach the problem as a multiclass or single-label classification task, and return the type with the highest probability.

%% --- Figure: 2 architectures
%
%

%

% -------
\subsection{Input Components}
\label{subsec:inputcomponents}

We consider three input components: \inputA, \inputB, and \inputC.  
Each of these input vector spaces aims to represent a particular kind of information associated with an entity.  

% --- Input A
Component \inputA is the main input representation, and consists of word embeddings of short entity descriptions.  
Specifically, for an entity $e$ we retrieve its short description $s_e$ in DBpedia.  
We then assign to each token $w$ in $s_e$ its 300-dimensional vector $\vect{v}_w$ in the \emph{word2vec} pre-trained word embeddings~\citep{MIKOLOV:2013:DRO}.  
Input \inputA is then simply the centroid $\vect{c}_e$ of these word embeddings for $e$.

% --- Input B
Component \inputB comprises the short descriptions of the entities that are related to the input entity $e$.  
Given $e$, we retrieve the set of related entities $R_e$, and obtain for each $e' \in R_e$ the centroid $\vect{c}_{e'}$ of word embeddings in its short description as before.  
We define \inputB as $avg_{e' \in R_e} \vect{c}_{e'}$, i.e., the centroid of these related entities' centroids.  

% --- Input C
Finally, component \inputC represents the frequency of the types of related entities.  
Formally, given an entity $e$ and its related entities $R_e$, the type frequency vector of related entities is defined as $(f_1, ..., f_n)$ where $f_i$ counts how many entities in $R_e$ are assigned to type $t_i$, for each of the $n$ labels in the universe of types.  

% --- "Plus" notation
We denote with ``+'' the fact that more than a single input component is provided to the model.  
For example, \inputA~+~\inputB means that \inputA and \inputB are provided, while \inputC is ignored.  

  % approach
\section{Experimental Setup}
\label{sec:setup}

In this section we present our experimental datasets, evaluation metrics, and parameter settings for the neural architectures.

% -------
\subsection{Datasets}
\label{subsec:datasets}

% --- Datasets overview
We create two test datasets, each consisting of 1,000 entities. \firstDataset represents established entities, while \secondDataset focuses on emerging entities.
Before we detail these two datasets, we describe the entity selection method that is shared by both.

% --- Pseudo-random drawing approach
The distribution of entity types in DBpedia is skewed, e.g., type \texttt{Person} is over-represented.  
We let $T_{top}$ be the set of all top-level types in the DBpedia Ontology (but removing \texttt{dbo:Agent}, and adding \texttt{dbo:Person} and \texttt{dbo:Organization} instead,  for informative purposes).  
We propose a pseudo-random approach for producing a test set of 1k instances, by drawing entities in a balanced way to represent as many types in $T_{top}$ as possible, with a minimum amount of $m = 10$ entities.  
For each type $t \in T_{top}$, we draw $m$ entities that are typed with $t$.  
If there exist less than $m$ such entities, we reserve one of those entities for training and draw the remaining ones for testing.  
This ensures that each type is observed at least once in training data.  
Finally, we draw the remaining needed entities randomly to end up with a total of 1k test instances.  

% --- Datasets in detail

\emph{\firstDataset} is generated by drawing entities that have type assignments (``instance types'') in DBpedia 2016-10, using the balanced pseudo-random approach previously described.  We apply the following additional constraints for entities: (i) they must have types predicted by SDTypes, to facilitate comparison, and (ii) they must have a short description in DBpedia. Each entity is then labeled with a single, most specific type in DBpedia instance types.  

\emph{\secondDataset} represents emerging entities and is created by picking entities with types in DBpedia Live,\footnote{Retrieved % on 06.03.18
through SPARQL endpoint \uri{http://dbpedia-live.openlinksw.com/sparql.}} such that these entities do not have types in DBpedia 2016-10.  
We also require that entities (i) have types predicted by SDTypes, and (ii) have a short abstract in DBpedia Live.
We use the pseudo-random approach previously described, but without the optional reservation of training instances.  

% --- Training set
The same training set is used in both cases, which is the universe of all DBpedia entities with types (3,047,794 in total), excluding those that are present in either of the test datasets. %\firstDataset or \secondDataset. 

% -------
\subsection{Evaluation metrics}
\label{subsec:evaluationmetrics}

% --- Overall eval. metric
We employ a rank-based evaluation for finding the most correct entity type.  
Since we are interested in predicting a single entity type, we use normalized Discounted Cumulative Gain at rank 1 (NDCG@1) as our evaluation metric.  
We consider different ways of computing gain according to a distance $d(t_a, t_g)$ between an assigned entity type $t_a$ and a ground truth type $t_g$.  
The motivation is to take into account the hierarchy of types~\citep{BALOG:2012:HTT}.
For example, predicting type \texttt{Person} for a correct type \texttt{Athlete} is a less severe error than predicting \texttt{Scientist}.  

% --- Strict
Firstly, a strict gain scoring is defined as follows.  
A score of 1 is given if $d(t_a, t_g) = 0$, otherwise the score is 0.  
Note that strict scoring is thus equivalent to classification accuracy.
%--- Lenient
Alternatively, a scoring method can reward close misses, e.g., when predicting wrongly an ancestor or descendant of the correct type.  
We implement this lenient metric using two different gain measures.  
A \emph{linear} decay function is defined by $G(t) = 1 - \frac{d(t_0, t_1)}{h}$, 
where $h$ is the depth of the ontology ($h = 6$ in DBpedia 2016-10).  
An \emph{exponential} decay function is instead defined as $G(t) = b^{-d(t_0, t_1)}$,
where $b$ is the base of the exponent (set to 2 in this paper).

% -------
\subsection{Parameter Settings}
\label{subsec:parametersettings}

% --- Params. optim. and other default settings
We optimize \complexArch with all three input components, by experimenting with several parameter settings as follows.  
We do a learning rate sweep search in $\{10^{-k} :  k \in {1, 2, 3}\}$, and a finer search in the interval [0.05, 0.5] with steps of 0.05.  
We try different optimizers, specifically, SGD, SGD with momentum, and Adam.
Furthermore, we test adding dropout in three network positions (before \texttt{hidden\_M.1}, after \texttt{hidden\_M.2}, between \texttt{hidden\_M.1} and \texttt{hidden\_M.2}),
each using probability $p \in \{0.2, 0.4\}$.
We set a hidden layer size of 512, with ReLU activation nodes, and use categorical cross entropy as the loss function.  

% --- Optim. results, and strategy across all models
We found the following parameter settings to perform best: SGD optimizer with learning rate of 0.1, and no dropout.  
We then train all the models for both architectures using these settings.  
For inputs different than \inputA~+~\inputB~+~\inputC, we ignore the missing input portion(s) from the model.  
For example, in \inputA~+~\inputB, we remove \texttt{input\_C} in Fig.~\ref{fig:neuralnetsimple}, and remove \texttt{input\_C}, \texttt{hidden\_C.1} and \texttt{hidden\_C.2} in Fig.~\ref{fig:neuralnetcomplex}.

% -------
\subsection{Retrieval Baseline}
\label{subsec:text_baseline}

Since we use entity descriptions as the main input component, the question naturally arises: How well would a traditional retrieval method perform on this task? 
The guiding observations is that for some entities, the type label occurs in the description, most likely in the copula relation \emph{``to be a''} with the entity name.  
An example is the label ``soccer player'' of the type \texttt{SoccerPlayer} in ``Alex Morgan is an American soccer player.''  
Hence, for a given entity, we score each type $t$ against the entity's description using the BM25 retrieval model.  Specifically, the camel-case DBpedia type label (\texttt{SoccerPlayer}) is converted into a lower-case phrase (``soccer player'') and used as a query.  We then take the top ranked type (i.e., the one with the highest BM25 score) as the prediction.  
  % expsetup
\section{Experimental Results}
\label{sec:results}

% --- Results table
\begin{table*}[t]
  \centering
  \vspace*{-0.75\baselineskip}
  \caption{Entity typing results, measured in terms of NDCG@1.  
		  $s$ is the standard deviation for the averaged performances of the neural models.  
		  \highSigBASE~ denotes a statistically significant difference w.r.t. SDType at $p<0.001$.
		  Statistical significance of each model in \complexArch versus the corresponding one in \simpleArch, at $p<0.05$ and $p<0.001$, is denoted by \sigARCHS~ and \highSigARCHS, respectively.
  }
  \vspace*{-0.75\baselineskip}
  \footnotesize  % *** NOTE
  \begin{tabular}{ l | l l | l l | l l || l l | l l | l l }
    \toprule
    \multirow{3}{*}{Model}
    & \multicolumn{6}{c||}{\firstDataset (Established entities)}
    & \multicolumn{6}{c}{\secondDataset (Emerging entities)} \\
     & \multicolumn{2}{c|}{Strict} & \multicolumn{2}{c|}{Linear} & \multicolumn{2}{c||}{Exponential}
     & \multicolumn{2}{c|}{Strict} & \multicolumn{2}{c|}{Linear} & \multicolumn{2}{c}{Exponential} \\
     & \metric & $s$ & \metric & $s$ & \metric & $s$
     & \metric & $s$ & \metric & $s$ & \metric & $s$ \\
    \midrule
    BM25
	& 0.2335 & - & 0.3211 & - & 0.2798 & - & 0.2076 & - & 0.3533 & - & 0.2826 & - \\
    \midrule
	SDType
	& 0.8020 & - & 0.8562 & - & 0.8331 & - & 0.6970 & - & 0.7873 & - & 0.7451 & - \\
	\midrule
    \multicolumn{5}{l}{\emph{\simpleArch}} \\
    \inputA
	& 0.8578\highSigBASE & 0.0025 & 0.8980\highSigBASE & 0.0028 & 0.8796\highSigBASE & 0.0028 & \textbf{0.7870}\highSigBASE & 0.0039 & \textbf{0.8617}\highSigBASE & 0.0045 & \textbf{0.8272}\highSigBASE & 0.0042 \\
	\inputB
	& 0.7722 & 0.0051 & 0.8028 & 0.0049 & 0.7895 & 0.0047 & 0.3664 & 0.0051 & 0.5381 & 0.0033 & 0.4419 & 0.0039 \\
	\inputC
	& 0.7222 & 0.0050 & 0.7571 & 0.0043 & 0.7414 & 0.0048 & 0.2950 & 0.0054 & 0.3781 & 0.0044 & 0.3341 & 0.0045 \\
	\inputA+~\inputB
	& \textbf{0.8864}\highSigBASE & 0.0055 & \textbf{0.9193}\highSigBASE & 0.0035 & \textbf{0.9045}\highSigBASE & 0.0043 & 0.7700\highSigBASE & 0.0049 & 0.8558\highSigBASE & 0.0042 & 0.8164\highSigBASE & 0.0040 \\
	\inputA+~\inputC
	& 0.8766\highSigBASE & 0.0057 & 0.9072\highSigBASE & 0.0066 & 0.8935\highSigBASE & 0.0063 & 0.7532\highSigBASE & 0.0092 & 0.8355\highSigBASE & 0.0074 & 0.7974\highSigBASE & 0.0080 \\
	\inputB+~\inputC
	& 0.7828 & 0.0098 & 0.8157 & 0.0098 & 0.8006 & 0.0096 & 0.3756 & 0.0181 & 0.5251 & 0.0718 & 0.4430 & 0.0406 \\
	\inputA+~\inputB+~\inputC
	& 0.8748\highSigBASE & 0.0090 & 0.9074\highSigBASE & 0.0091 & 0.8930\highSigBASE & 0.0091 & 0.7462\highSigBASE & 0.0058 & 0.8354\highSigBASE & 0.0030 & 0.7944\highSigBASE & 0.0035 \\ \midrule
    \multicolumn{5}{l}{\emph{\complexArch}} \\
    \inputA
	& 0.8558\highSigBASE & 0.0029 & 0.8956\highSigBASE & 0.0028 & 0.8777\highSigBASE & 0.0026 & \textbf{0.7816}\highSigBASE & 0.0050 & 0.8587\highSigBASE & 0.0057 & \textbf{0.8230}\highSigBASE & 0.0056 \\
	\inputB
    & 0.7788 & 0.0050 & 0.8070 & 0.0050 & 0.7947 & 0.0049 & 0.3696 & 0.0130 & 0.5453 & 0.0116 & 0.4475 & 0.0125 \\
	\inputC
    & 0.7224 & 0.0060 & 0.7586 & 0.0066 & 0.7424 & 0.0061 & 0.2854 & 0.0060 & 0.3690 & 0.0055 & 0.3244 & 0.0055 \\
	\inputA+~\inputB
	& 0.8896\highSigBASE & 0.0026 & 0.9219\highSigBASE & 0.0030 & 0.9074\highSigBASE & 0.0025 & 0.7766\highSigBASE & 0.0057 & \textbf{0.8600}\highSigBASE & 0.0041 & 0.8216\highSigBASE & 0.0046 \\
	\inputA+~\inputC
	& 0.8926\highSigBASE\highSigARCHS & 0.0046 & 0.9256\highSigBASE\highSigARCHS & 0.0034 & 0.9108\highSigBASE\highSigARCHS & 0.0034 & 0.7670\highSigBASE\sigARCHS & 0.0068 & 0.8490\highSigBASE\sigARCHS & 0.0047 & 0.8107\highSigBASE\highSigARCHS & 0.0055 \\
	\inputB+~\inputC
	& 0.8134\highSigARCHS & 0.0068 & 0.8431\highSigARCHS & 0.0068 & 0.8299\highSigARCHS & 0.0068 & 0.3802 & 0.0242 & 0.5286 & 0.0777 & 0.4470 & 0.0464 \\
	\inputA+~\inputB+~\inputC
	& \textbf{0.8958}\highSigBASE\highSigARCHS & 0.0027 & \textbf{0.9284}\highSigBASE\highSigARCHS & 0.0033 & \textbf{0.9138}\highSigBASE\highSigARCHS & 0.0026 & 0.7556\highSigBASE & 0.0108 & 0.8487\highSigBASE\highSigARCHS & 0.0076 & 0.8056\highSigBASE\highSigARCHS & 0.0092 \\
 \bottomrule
  \end{tabular}
  \normalsize  % *** NOTE
  \label{table:results}
  \vspace*{-0.5\baselineskip}
\end{table*}

%

% --- RQs
With our experiments, we seek to answer the following three research questions: (RQ1) Can a neural approach, using only entity descriptions, outperform the current state of the art (SDType), which is based on heuristic link-based type inference?; (RQ2) Can entity relationship information contribute to type prediction performance?; (RQ3) Which of the two proposed neural architectures (\simpleArch vs. \complexArch) performs better?

% --- Results table, notation, significances

Table~\ref{table:results} presents the results.  
We evaluate both \simpleArch and \complexArch using all combinations of input components.  
Scores reported for each neural model are averaged from 5 independent training sessions.  
In each session, a model is trained for a maximum of 50 epochs, with early stopping implemented in order to prevent overfitting.  
Early stopping is configured to stop training when no improvement is observed for 5 epochs.  
Using respective two-tailed paired t-tests, we assess statistical significance (i) against the SDType baseline, and (ii) of each model in \complexArch versus the corresponding one in \simpleArch.

As the results clearly indicate, the type prediction task is much more involved than plain text retrieval.  
The BM25 ranker is inferior to all the other methods.  
Henceforth, we will be focusing on the SDType as a baseline.

% --- Analysis
% - RQ1
\emph{RQ1.} For answering our first research question, we compare input configuration \inputA in both \simpleArch and \complexArch against the baseline method SDType.  
In both architectures, it is clear that the neural approach using short entity descriptions is able to significantly outperform the baseline in both \firstDataset and \secondDataset across all evaluation measures. 

We are are interested in comparing each method's ability to predict types in a given top-level branch (esp. given that types inferred by SDType are known to be often generic and high up in the type hierarchy~\citep{PAULHEIM:2013:TIO}).  
Our simple architecture \simpleArch outperforms SDType substantially in most of the most prominent types, such as \texttt{Person}, \texttt{Place} and \texttt{Organization} ($58.8\%$ of \firstDataset among these three types).  
On the other hand, SDType has better performance on much less prominent top-level types, such as \texttt{Biomolecule}, \texttt{TopicalConcept}, and \texttt{Language}.  
Given that these account for $1.4\%$ or less in \firstDataset, this could be attributed to a lack of training data for these top-level types.

% - RQ2
\emph{RQ2.} Our second research question considers the effect of adding optional inputs \inputB and \inputC.  
Observing results on \firstDataset, performances improve in both architectures when including optional inputs.  
Specifically, in \simpleArch, using inputs \inputA~+~\inputB is the best performing model, thus proving that the neural transformations properly capture both components of similar structure (i.e., centroids of word embeddings).  
In \complexArch, configuration \inputA~+~\inputB~+~\inputC has highest performance, as another evidence of the benefits of representations of related entities.  
When comparing \simpleArch and \complexArch, the additional hidden layers per input reward the optional inputs significantly. 
In evaluating the addition of optional inputs on \secondDataset, we see that these actually deteriorate performance compared to using only short entity descriptions.  
Recall that this dataset represents emerging entities, and therefore an entity might not have enough relationships compared to entities in \firstDataset.  
Consequently, inputs \inputB and \inputC are much more sparse and do not contain the same rich data as in \firstDataset. 

% - RQ3
\emph{RQ3.} To answer the final research question, we compare \simpleArch to \complexArch.  
In \firstDataset, input \inputA~+~\inputB~+~\inputC in \complexArch has a slightly better score than \inputA~+~\inputB in \simpleArch.  
It is also interesting to note that \complexArch almost scores just as high using only \inputA~+~\inputC.  
On the other hand, when comparing on \secondDataset, it is clear that short entity descriptions from \inputA are more valuable in the case of sparse relationship data than \inputB and \inputC together.  
Here, \simpleArch provides mostly almost identical scores to \complexArch, and thus \simpleArch is preferable when considering time and resources required for training the model.  
The only configurations performing worse than the baseline, are those without input \inputA, which further confirms the importance of short descriptions of the entities themselves.  

  % results
\section{Conclusions}
\label{sec:concl}

We have addressed the problem of automatically assigning a type to a given entity in a knowledge base, proposed two simple neural network architectures, and experimented with a variety of input entity representations.  
A main finding is that even these simple neural approaches, relying on limited input, are able to provide a significant improvement over the existing state-of-the-art, which requires semantically rich information as input.  
In future work, we would like to evaluate our approach on other KBs and explore alternative network architectures and input representations.

  % concl

% Bibliography at the end, not splitted with figures inside
\FloatBarrier  % from {placeins} package

\bibliographystyle{ACM-Reference-Format}
\bibliography{ictir2019-entity-typing}

\end{document}